\documentclass[a4paper,11pt]{article}
\pdfoutput=1 
\usepackage{jheppub} 

\usepackage[utf8]{inputenc}

\usepackage{color}      % Non-standard graphics support
\graphicspath{{figs/}}

\usepackage{amsfonts}       % AMS fonts
\usepackage{amsbsy}     	 % Bolds (vectors, tensors, etc.)

\newcommand{\sina}{\sin\alpha}
\newcommand{\cosa}{\cos\alpha}

\newcommand{\sa}{s_{\alpha}}
\newcommand{\ca}{c_{\alpha}}
\newcommand{\ztwo}{\mathbb{Z}_2}
\newcommand{\half}{\frac{1}{2}}
\newcommand{\hp}{{H^\pm}}

\newcommand{\mhp}{{m_{H^\pm}}}

\newcommand{\M}{{\cal M}}

\newcommand{\h}{{\cal H}}
\newcommand{\ii}{\mathrm{i}}
\newcommand{\BR}{\text{BR}}

\newcommand*{\THDMC}{{\sc 2hdmc}}

\newcommand{\HiB}{{\sc HiggsBounds}}

\newcommand{\reffig}[1]{{figure~\ref{#1}}}
\newcommand{\refeq}[1]{{eq.~(\ref{#1})}}
\newcommand{\eq}[1]{{(\ref{#1})}}

\newcommand{\be}{\begin{equation}}
\newcommand{\ee}{\end{equation}}

 \preprint{LU-TP 13-12}

\title{Higgs properties in a broken Inert Doublet Model}

\author[a]{Rikard Enberg,}
\author[b]{Johan Rathsman,}
\author[a]{and Glenn Wouda}

\affiliation[a]{Department of Physics and Astronomy,
 Uppsala University, Box 516, SE--751 20 Uppsala, Sweden}

\affiliation[b]{Department of Astronomy
 and Theoretical Physics,
 Lund University, SE--223 62 Lund, Sweden}

\emailAdd{Rikard.Enberg@physics.uu.se}
\emailAdd{Johan.Rathsman@thep.lu.se}
\emailAdd{Glenn.Wouda@physics.uu.se}

\abstract{We consider a model for the Higgs sector with two
scalar doublets and a broken $\ztwo$ symmetry, the Stealth Doublet Model, where the $\ztwo$ symmetry is manifest in the Yukawa sector but broken by the scalar potential. 
This model can be seen as a
generalization of the Inert Doublet Model.  One of the doublets is the
Higgs doublet that participates in electroweak symmetry breaking and couples to
fermions. The other doublet does not couple to fermions at tree level and does
not acquire a vacuum expectation value. 
The broken $\ztwo$ symmetry leads to interesting phenomenology such as mixing between the two
doublets and charged and CP-odd scalars that can be light and have unusual decay channels.
We present theoretical and experimental constraints on the model
and consider the recent observation of a
Higgs boson at the LHC. The data on the $H\to\gamma\gamma$
channel can be naturally accommodated in the model, with either the lightest or
the heaviest CP-even scalar playing the role of the observed particle.

\vspace{5cm}
~
}

\begin{document}
\notoc
\maketitle
\flushbottom

 \section{Introduction}

 The ATLAS \cite{ATLAShiggs} and CMS \cite{CMShiggs} experiments at the Large
 Hadron Collider (LHC) have discovered a new particle that exhibits all the
 features of a Higgs boson (for the most recent data see e.g.\
 \cite{ATLAS-CONF-2013-012,ATLAS-CONF-2013-013,CMS-PAS-HIG-13-001}). It will require hard work to
 uncover if this is the Standard Model (SM) Higgs boson or not, but the data
 seem to be compatible with the SM. If it is a Higgs boson, but not the one
 predicted by the SM, then there is likely an extended Higgs sector with
 additional Higgs bosons.

 Arguably, the most ``standard'' Higgs scenarios beyond the SM are the minimal
 supersymmetric Standard Model (MSSM), general two-Higgs doublet models (2HDM),
 or perhaps the next-to-minimal supersymmetric standard model (NMSSM); see in
 particular ref.~\cite{Branco:2011iw} for a recent review of 2HDMs. These models
 all predict a similar set of additional Higgs bosons, with associated
 dominating production and decay channels, and most searches are devoted to
 these channels.

 There is a real possibility that Nature is not described by one of these
 standard scenarios, and if it is not, the standard searches may not be
 appropriate. It is therefore important to consider alternative scenarios. One
 such alternative is the Inert Doublet Model (IDM)
 \cite{Deshpande:1977rw,Ma:2006km,Barbieri:2006dq}, where there is a conserved
 $\ztwo$ parity, such that while one scalar plays the role of a SM-like Higgs
 boson, the other scalars do not couple to fermions. In
 particular, the lightest scalar is a stable dark matter candidate.

 In this paper we present the Stealth Doublet Model (SDM), a generalization of
 the IDM where the $\ztwo$ symmetry is broken, and discuss its impact on
 Higgs physics at the LHC. In particular we study the SDM in relation to the LHC
discovery and exclusion results. When the $\ztwo$ symmetry of the IDM is
 broken, the lightest scalar is not stable, and couplings of the fermiophobic
 particles to fermions are generated at one-loop level. The resulting
 phenomenology is very different from the standard scenarios discussed above,
 and as we will show below is also compatible with the discovered particle at
 LHC.

 The $\ztwo$ symmetry is manifest in the Yukawa sector but broken in the scalar
 potential. We consider this as a way to parametrize our ignorance of the
 symmetry breaking mechanism, in a similar way to how supersymmetry breaking is
 parametrized in e.g.\ the MSSM.

 The particle content of the SDM is the same as in CP-conserving 2HDMs:\ there
 are two CP-even neutral scalars, $h^0$ and $H^0$, one of which should be the
 Higgs boson discovered at the LHC, a CP-odd neutral scalar $A^0$, and a charged
 scalar $\hp$. The interactions of these particles are quite different from
 those of 2HDMs, however. 

In our model, the $A^0$ and $\hp$ have no tree-level
couplings to fermions:\ they are \textit{fermiophobic}. All their couplings to fermions
are instead generated at the one-loop level. Their dominating
 production and decay modes can therefore be different than in the standard scenarios.
 Consequently, $A^0$ and $\hp$ can be lighter in the SDM than in standard
 scenarios, since flavor constraints and LEP limits do not apply.
For example, if the charged scalar is the lightest scalar, its main decay is typically $\hp\to
 W^\pm\gamma$. In addition, electroweak precision tests (EWPT) allow both
 lighter or heavier $h^0,H^0$ in our model than in the SM.

 In the rest of this paper we describe the SDM and its parameters. We
 consider the various constraints on the model and the observed Higgs boson signal at LHC.
Finally we briefly consider the charged  scalar and its
 decay and production channels. We
 leave detailed analyses of the model for upcoming papers~\cite{upcoming}, where
 we will also compute all decay widths of the fermiophobic scalars and discuss
 LHC phenomenology in more detail. A preliminary presentation of this model can be found in \cite{Wouda:2010zz}.

 \section{The Stealth Doublet Model}\label{sect:model}

 In brief, the model consists of adding another doublet to the Standard Model
 scalar sector, and introducing a discrete $\ztwo$ parity between the two
 doublets in the Yukawa sector. This parity is broken in the scalar potential. The scalar potential is thus the potential
 of two-Higgs doublet models (2HDM) with two hypercharge $Y=1$ scalar doublet
 fields $\Phi_i=(\phi_i^+,\phi_i^0)^T$,
\begin{align}
V &= m_{11}^2\Phi_1^\dagger\Phi_1+m_{22}^2\Phi_2^\dagger\Phi_2
-[m_{12}^2\Phi_1^\dagger\Phi_2+{\rm h.c.}]\nonumber\\
& +\half\lambda_1(\Phi_1^\dagger\Phi_1)^2
+\half\lambda_2(\Phi_2^\dagger\Phi_2)^2
+\lambda_3(\Phi_1^\dagger\Phi_1)(\Phi_2^\dagger\Phi_2)
+\lambda_4(\Phi_1^\dagger\Phi_2)(\Phi_2^\dagger\Phi_1)
\nonumber\\
& +\left\{\half\lambda_5(\Phi_1^\dagger\Phi_2)^2
+\big[\lambda_6(\Phi_1^\dagger\Phi_1)
+\lambda_7(\Phi_2^\dagger\Phi_2)\big]
\Phi_1^\dagger\Phi_2+{\rm h.c.}\right\}\,,
\label{eq:Vpotential}
\end{align}
where we only consider CP-conserving models and therefore take all parameters to be real.
 The $\ztwo$ symmetry transformation can be taken as
 $\Phi_1\to\Phi_1$ and $\Phi_2\to -\Phi_2$. This parity is broken by the last
 three ``hard-breaking'' terms in \refeq{eq:Vpotential} and by the
soft-breaking term $m_{12}^2\Phi_1^\dagger\Phi_2+{\rm h.c.}$

 Furthermore, in this model the second doublet does not get a vacuum expectation
 value (vev) and is therefore not really a Higgs doublet. However, because the
 $\ztwo$ symmetry is broken the two doublets can mix. Studying a model where the
 vev resides solely in one of the doublets is equivalent to working in the Higgs basis (see e.g.\
 \cite{Davidson:2005cw,Haber:2006ue,Branco:2011iw}). 
 The Higgs basis is often useful for
 analyzing a general two-Higgs doublet model, but in our model, this is
 the physical basis, with $v=v_1\approx 246$~GeV. Similarly to the IDM, the SDM does not have a parameter
 $\tan\beta=v_2/v_1$, and
 cannot be obtained by simply taking the limit $\tan\beta\to 0$ or
 $\tan\beta\to\infty$.

Minimizing the potential yields the conditions
\begin{align}
 m_{11}^2 =  -\half v^2 \lambda_1, \qquad m_{12}^2 = \half v^2 \lambda_6,
 \label{eq:minim}
\end{align}
and $m_{22}^2$ is thus a free parameter.
Further constraints on the parameters will be discussed below.

 Generally in two-Higgs doublet models where CP is conserved, there are two
 CP-even neutral scalars, one CP-odd neutral scalar $A^0$, and a charged scalar
 $H^\pm$. The two doublets can be written in unitary gauge as
\begin{align}
\Phi_1 =  \begin{pmatrix} 0 \\ \dfrac{v+\phi_1^0}{\sqrt{2}} \end{pmatrix} , \qquad
\Phi_2 =  \begin{pmatrix} H^+ \\ \dfrac{\phi_2^0 + \ii A^0}{\sqrt{2}} \end{pmatrix},
\end{align}
where $\phi_{1,2}^0$
 are the neutral CP-even interaction eigenstates, whose mass matrix is not
 diagonal:
\be \label{eq:massmatrix}
\M^2 = \begin{pmatrix}
\lambda_1 v^2 & \lambda_6 v^2 \\
\lambda_6 v^2 \, & \, m_{22}^2 + \lambda_{345} v^2 \\
\end{pmatrix}
 =
\begin{pmatrix}
\lambda_1 v^2 & \lambda_6 v^2 \\
\lambda_6 v^2 \, & \, m_A^2 + \lambda_5 v^2 \\
\end{pmatrix},
\ee where
 $\lambda_{345}=\lambda_3+\lambda_4+\lambda_5$. The physical CP-even
 mass eigenstates are given by
\begin{align}
H^0 &=  \phi_1^0 \cosa + \phi_2^0 \sina\\
h^0 &= - \phi_1^0 \sina + \phi_2^0 \cosa,
\end{align}
where the mixing angle $\alpha$ is obtained by diagonalizing the matrix $\M^2$. The couplings of $h^0$ and $H^0$ to $Z^0 Z^0$ and $W^\pm W^\mp$ are thus suppressed compared to the SM Higgs by factors $\sina$ and $\cosa$, respectively. Taking $H^0$ to be the heavier state, the masses of $h^0,H^0$ are
\begin{align}
 m_h^2 & = c_\alpha^2 m_A^2 + s_\alpha^2 v^2 \lambda_1 +c_\alpha^2 v^2 \lambda_5
 - 2 s_\alpha c_\alpha v^2 \lambda_6 \label{eq:Hhmasses1}\\ m_H^2 & = s_\alpha^2
 m_A^2 + c_\alpha^2 v^2 \lambda_1 + s_\alpha^2 v^2 \lambda_5 + 2 s_\alpha
 c_\alpha v^2 \lambda_6, \label{eq:Hhmasses2}
\end{align}
where we defined the abbreviations $\sa\equiv\sina, \, \ca\equiv\cosa$. 
The masses of the remaining states are 
\begin{align}
 &m_A^2   = m_\hp^2 - \half v^2 (\lambda_5-\lambda_4) \label{eq:Amass}\\
&m_\hp^2 = m_{22}^2 + \half v^2 \lambda_3. \label{eq:Hpmass}
\end{align}
One may solve eqs.\ (\ref{eq:Hhmasses1}--\ref{eq:Hpmass}) for the
 parameters $\lambda_{1,3,4,5}$ so that the masses of the scalars can be used as
 model parameters. The mixing angle is given by
\be
\sin 2\alpha =
 \frac{2v^2\lambda_6}{m_H^2-m_h^2}.
\label{eq:s2a}
\ee
From eqs.\ \eq{eq:massmatrix} and \eq{eq:s2a}, we see that when the $\ztwo$ parity is
 exact (i.e.\ $\lambda_6=0, m_{12}^2=0$), there is no mixing of the CP-even states and we
 recover the Inert Doublet Model. In fact, all our results reduce to the IDM
when letting $\lambda_6\to 0$ and either $\sina\to 1$, $\cosa\to 0$ or $\sina\to
0$, $\cosa\to 1$. In this sense, our model is a generalization of the IDM.

 The $\ztwo$ symmetry, if exact, would forbid flavor-changing neutral currents (FCNC).
 If the symmetry is broken, there could potentially be large FCNC. In this model the symmetry
is broken in the potential and FCNC do not occur on the tree-level, but only at
 two-loop level \cite{upcoming}.

 The scalar potential \eq{eq:Vpotential} has ten free parameters when requiring all
 couplings to be real. Two are eliminated by the minimization conditions
 (\ref{eq:minim}). 
 We will choose the masses of the scalar states $m_h, \
 m_H, \ m_A$ and $m_{H^\pm}$ as four of the remaining parameters. To specify the
 amount of $\ztwo$ breaking, we choose to use $\sina$, 
which is related to $\lambda_6$ through \refeq{eq:s2a}. Three more
 parameters need to be specified. We take these to be $\lambda_2$, $\lambda_3$ and
 $\lambda_7$, since $\lambda_3$ determines the coupling between $h^0/H^0$ and
 $H^\pm$. In summary, we will use as the parameters of the model
\be
m_h, \ m_H, \ m_A, \ m_{H^\pm},  \ \sina , \ \lambda_2 , \ \lambda_3, \ \lambda_7 .
 \label{modelparams}
\ee

In the following we will usually choose
$\lambda_7=\lambda_6$ and $\lambda_2=\lambda_1$. 
 In our numerical calculations we will often choose $\lambda_3$
 to take the values $\lambda_3 = 0$, $2m_{\hp}^2/v^2$ and $4m_{\hp}^2/v^2$,
 corresponding to $m_{22}^2 = m^2_\hp$, 0  and $ -m^2_\hp$, respectively. We
 will also vary $\lambda_3$ and $\lambda_7$, within theoretically allowed
 regions, to deduce their impact on the signal strengths for $h^0/H^0
 \rightarrow \gamma \gamma$. 
 
 As a final step we must specify the Yukawa couplings of the model. The most
 general Yukawa Lagrangian in the Higgs Basis reads, in terms of the fermion
 flavor eigenstate \cite{Haber:2006ue}
\begin{align}
 -\mathcal{L}_{\text{Yuk}} &= \kappa^L_0 \bar{L}_L \Phi_1 E_R + \kappa^U_0
 \bar{Q}_L \widetilde\Phi_1\, U_R   +\kappa^D_0 \bar{Q}_L \Phi_1 D_R   \nonumber \\
  &+ \rho^L_0 \bar{L}_L \Phi_2 E_R + \rho^U_0 \bar{Q}_L
 \widetilde\Phi_2\, U_R  +\rho^D_0 \bar{Q}_L \Phi_2 D_R
 \label{Lyuk}
 \end{align}
where $\widetilde\Phi_i = -\ii \sigma_2 \Phi_i^*$. In
 order to obtain mass eigenstates, the matrix $\kappa^F_0$ can be diagonalized
by a biunitary transformation to obtain the diagonal mass matrix $M^F$ for fermions $F=U,D,L$.
 The correspondingly transformed $\rho^F$ matrix will in general be 
non-diagonal and will generate FCNC. To
 avoid FCNC at tree level, we impose positive $\ztwo$ parities on the fermions.
 This enforces $\rho^F = 0$ at tree level. The doublet $\Phi_2$ is therefore
 fermiophobic, and the states $H^\pm$, $A^0$, and $\phi_2^0$ do not interact with
 fermions at tree level. Fermions acquire mass through Yukawa couplings with the
 Higgs doublet $ \Phi_1 $ only. In this sense, our model is a Type I 2HDM.  The
 Higgs Yukawa Lagrangian is then
\be
-\mathcal{L}_{\text{Yuk}}
 =\dfrac{m_f}{v} \bar\psi_f \psi_f \left( \ca \,  H  -  \sa \,  h\right) 
\ee
for all fermions $f$. 
The $\ztwo$ breaking terms will lead to
 couplings between $\Phi_2$ and fermions at one-loop level. The resulting
 $\rho^F$ matrices are  diagonal at one-loop level \cite{upcoming}. At higher
 orders, $\rho^F $ will have off-diagonal elements and introduce FCNC,
  but these will be two-loop suppressed \cite{upcoming}.
In general one could consider constraints from the renormalization group evolution of the Yukawa couplings
along the lines of Ref.~\cite{Bijnens:2011gd}.

 \section{Constraints on the model}\label{sect:constraints}

 Let us now discuss the constraints that apply to the model. These are
 analyzed using the two-Higgs doublet model calculator
 \THDMC~\cite{Eriksson:2009ws,Eriksson:2010zzb}, where we have implemented the
 SDM as a new model. This makes it straightforward to obtain theoretical
 constraints, oblique parameters, branching ratios\footnote{Some of the
 important decay channels of $\hp$ and $A^0$ are one-loop processes that are not included in \THDMC, 
and will be briefly discussed below. They will be calculated and further discussed in
 \protect\cite{upcoming}.}, and cross sections.

 The parameters of the potential are constrained by demanding that the potential
 be bounded from below \cite{Deshpande:1977rw,Sher:1988mj} and that it respects perturbativity and
 tree-level unitarity\cite{Huffel:1980sk,Maalampi:1991fb,Kanemura:1993hm,Akeroyd:2000wc}.
 We will refer to these conditions as ``theoretical constraints.'' We do not
 list their explicit expressions here (see e.g.\ ref.~\cite{Eriksson:2009ws}),
 but will take them into account in all our calculations by using \THDMC. For details, see
 ref.~\cite{upcoming}.

 \begin{figure}[t]
\begin{centering}
 \begin{tabular}{cc}
 \includegraphics[width=0.45\columnwidth]{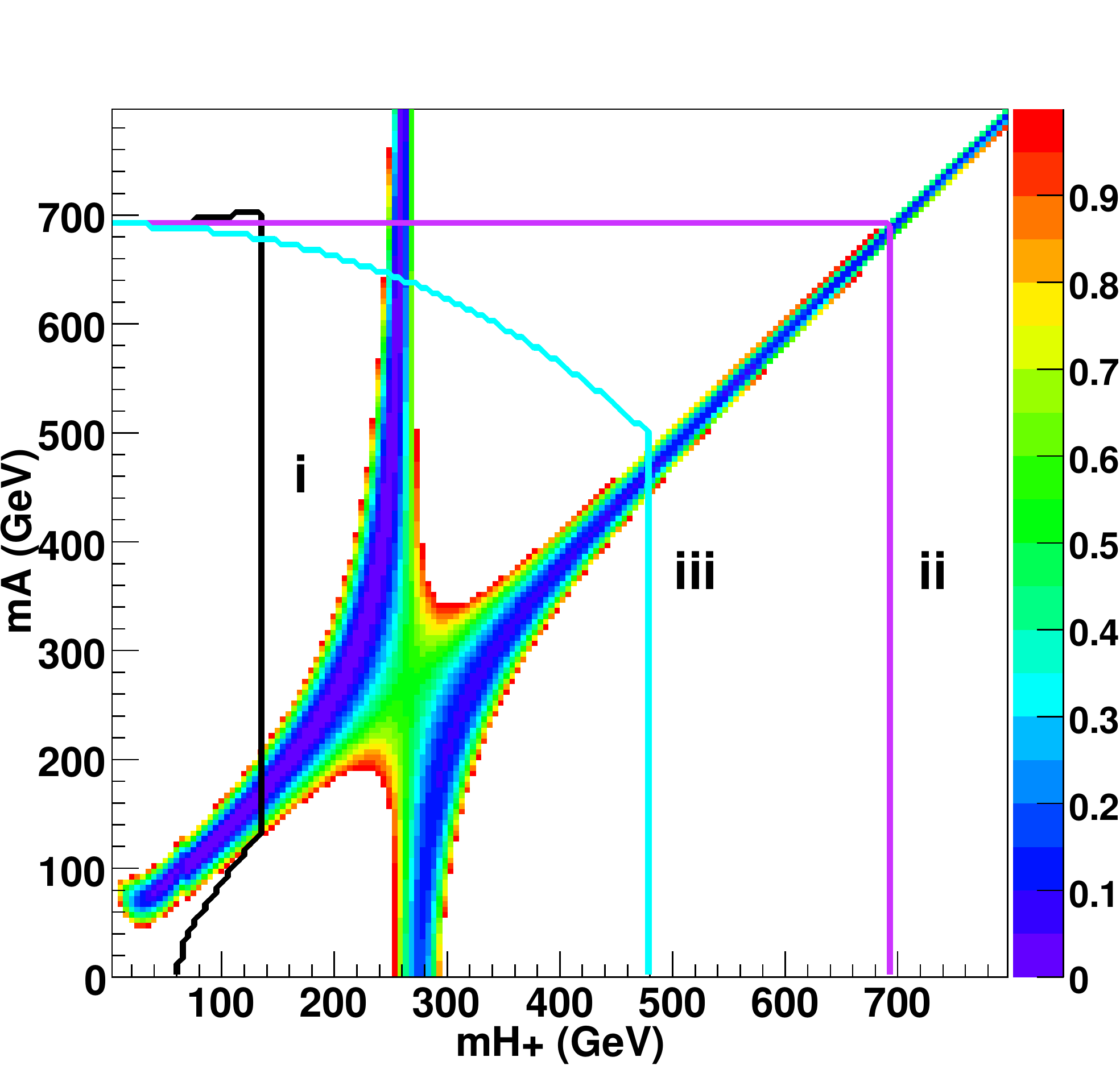}  &
 \includegraphics[width=0.45\columnwidth]{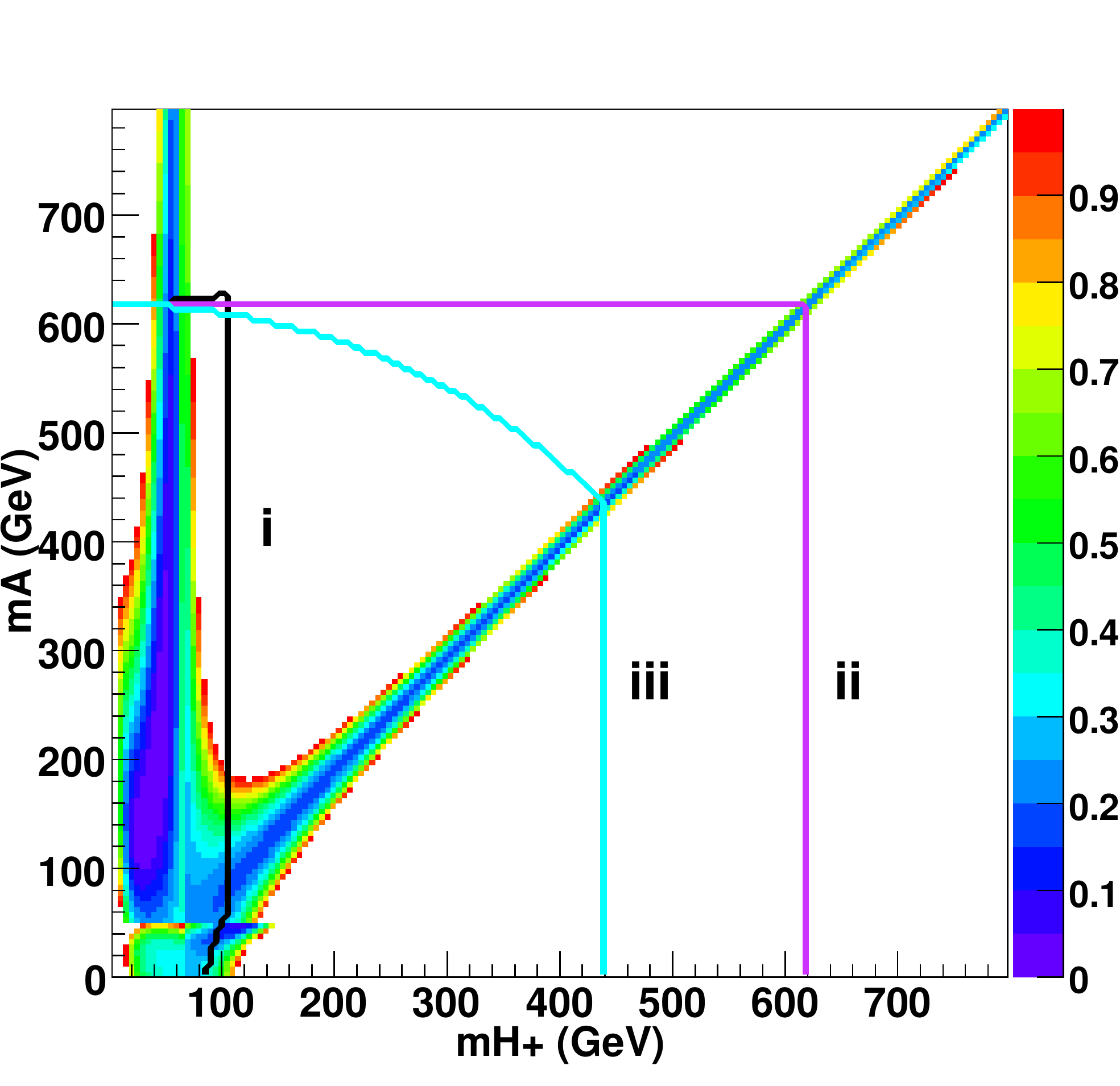} \\ 
 $m_h = 125 $ GeV,
 $\, m_H = 300$ GeV,  $s_\alpha = 0.9 $.   & $\quad$  $m_h = 75$ GeV, $m_{H} =
 125$ GeV,  $s_\alpha = 0.1 $.\\ (a) & $\quad$  (b)
\end{tabular}
 \caption{Examples of allowed regions in parameter space taking into account
 theoretical constraints and $S$ and $T$ values as a function of
  $m_{H^\pm}$ and $m_A$. The
 color displays the deviation from the center of the 90\% C.L.\ ellipse
 of Figure~10.7 in \cite{PhysRevD.86.010001}, taking the value 1 if on the limiting
 ellipse. White regions are outside the ellipse. The regions inside the 
black (i) ($m_{22}^2 = m^2_\hp
 \Rightarrow \lambda_3=0$), magenta (ii) ($m_{22}^2 = 0 \Rightarrow
 \lambda_3=2m_\hp^2/v^2$) and cyan (iii)  ($m_{22}^2 = -m^2_\hp \Rightarrow
 \lambda_3=4m_\hp^2/v^2$) lines fulfill the theoretical
 constraints for a given $m_{22}^2$ or $\lambda_3$-value. In this plot,
 $\lambda_2 = \lambda_1$ and $\lambda_7 = \lambda_6$. } \label{fig:constraints}
\end{centering}
\end{figure}

 The next set of constraints is given by electroweak precision tests (EWPT),
 most importantly those constraints imposed by loop contributions from the
 scalars to the gauge boson vacuum polarizations. Such corrections can be
 parametrized by the oblique parameters $S$, $T$ and $U$~\cite{Peskin:1991sw}.
 These do not depend explicitly on the potential parameters, but do
 so implicitly through the masses of the scalar particles in the model. All the
 scalars contribute to the oblique parameters, and the resulting expressions are
 lengthy but not particularly illuminating so we do not list them here. We use
  \THDMC{} to evaluate them and require the obtained values of $S$
 and $T$ to fall within the 90\% C.L.\ ellipse in Figure~10.7 of
 \cite{PhysRevD.86.010001}.

 Finally, there are constraints from collider searches at LEP, the Tevatron and
 the LHC. Constraints from LEP and the Tevatron and the 7 TeV LHC constraints 
(as implemented in the latest version of \HiB, v3.8)
will in the following be  included through the use of the \HiB{}
 program~\cite{Bechtle:2008jh,Bechtle:2011sb,Bechtle:2013gu} linked to \THDMC.
 These will be further discussed in the next section, but since the $A^0$ and
 $\hp$ in our model do not have the same interactions as in general 2HDMs, they are not
 much constrained by previous searches. The interactions of the CP-even scalar that has not
 been observed at the LHC, however, include the interactions of the SM Higgs
 boson, and it could therefore in principle have showed up in LHC searches.

 \section{Higgs and the LHC}

 Let us refer to the particle discovered by ATLAS and CMS as $\h$ and to the
 Higgs boson of the SM as $H_\text{SM}$. In our model, the $\h$ particle with
 mass roughly 125~GeV can be either the lighter $h^0$, with a heavier $H^0$ remaining to
 be discovered, or $\h$ can be the heavier $H^0$, while the $h^0$ is lighter and
 was not discovered at LEP or the Tevatron due to small couplings. We will refer
 to these two cases as Case~1 and Case~2, respectively.

 As we shall see below, if we want one of our CP-even scalars to account for the
 recent ATLAS data on the $\h\to\gamma\gamma$ decay channel, we must take the
 mixing $\sina$ relatively large in Case~1 and relatively small in Case~2. In
 \reffig{fig:constraints} we therefore present examples of regions in the
 $(m_\hp, m_A)$-plane that are allowed by theoretical and electroweak
 constraints for Case~1 and Case~2. For Case~1 we choose $\sa=0.9$, and for Case
 2, $\sa=0.1$. We further use $\lambda_2 = \lambda_1$ and $\lambda_7 =
 \lambda_6$. Note that since the
 $S$ and $T$ parameters do not explicitly depend on the $\lambda_i$, only some
 combinations of these parameters (i.e.\ the couplings and the masses of the
 scalars) are constrained by the electroweak precision data. We therefore
 additionally show in \reffig{fig:constraints} boundaries of regions admitted by
 the theoretical constraints for three different values of $\lambda_3$.

 From \reffig{fig:constraints} it is clear that the masses of the scalars should
 not exceed roughly 700~GeV in order to fulfill the theoretical constraints. 
Moreover, the largest allowed region is obtained for $m_{22}^2\approx 0$.
In order to give small enough contributions to the $S$ and $T$ parameters, $m_\hp$
 and $m_A$ must satisfy some approximate custodial symmetries:\ Define
 $M^2=m^2_H \sin^2\alpha + m^2_h\cos^2 \alpha$~\cite{Mahmoudi:2009zx}. Then 
 $m_A \approx m_\hp + 50$~GeV when $m^2_\hp \lesssim M^2$ or $m_A \approx m_\hp$ 
when $m^2_\hp \gtrsim M^2$ are allowed, respectively. When $
 m^2_\hp \approx M^2$, then $ 0 \lesssim m_A \lesssim 700 $ GeV is allowed.
These conclusions are similar to what was found in Ref.~\cite{Mahmoudi:2009zx}.

 In \reffig{fig:constraints} we have not yet included collider constraints. Let
 us estimate the LHC constraints on Case~1. Then $\h=h^0$, while $H^0$ is
 heavier. It is constrained by the LHC searches for the SM Higgs mainly through
 its decays $H^0\to ZZ^{(*)}\to 4\ell$, $H^0\to WW^{(*)}$ and
 $H^0\to\gamma\gamma$. Events where the $H^0$ is produced and decays in one of
 these channels will look exactly like the corresponding events for a SM Higgs
 at the same mass, so all experimental acceptances are identical. We can
 therefore compute the $\sigma \times \text{BR}$ for the relevant channels and
 use the LHC results to constrain the production of $H^0$.

 The most sensitive exclusion is production via $gg\to H^0$ and decay via
 $H^0\to ZZ$. Let us consider the most recent ATLAS exclusion in this channel
 shown in Figure~12a of ref.\ \cite{ATLAS-CONF-2013-013}. The SM curve in this
 plot shows  $\sigma\times\text{BR}$ for
$gg\to H_\text{SM}\to ZZ\to 4\ell$.
We want to rescale this curve to our model. The
 production processes are the same as in the SM, but the Yukawa coupling of
 $H^0$ to the heavy quark in the loop is suppressed by a factor $\ca$, and the
 same factor $\ca$ applies to vector--boson fusion and associated production.
 The decay vertex $H^0\to ZZ$ is also suppressed by the same factor $\ca$.
 However, the dominating decays to SM particles that contribute to the width
$\Gamma_{H^0}$ are $H^0\to ZZ, WW, b\bar b, t\bar t$. Again,
 all of these processes are suppressed by the same factor $\ca$ relative to the
 SM. In addition, there may be decays to the new scalars, $H^0\to h^0 h^0, H^+
 H^-,A^0 A^0, A^0 Z, \hp W^\mp$, if these channels are open. We may thus get a
 conservative upper bound on the branching ratio $\BR(H^0\to ZZ)$ by considering the case where
 none of the latter channels are open. The branching ratio is then the same as
 in the SM, but if any of the new channels open, it becomes smaller.

\begin{figure}[tb]
\begin{center}
\includegraphics[width=0.6\columnwidth]{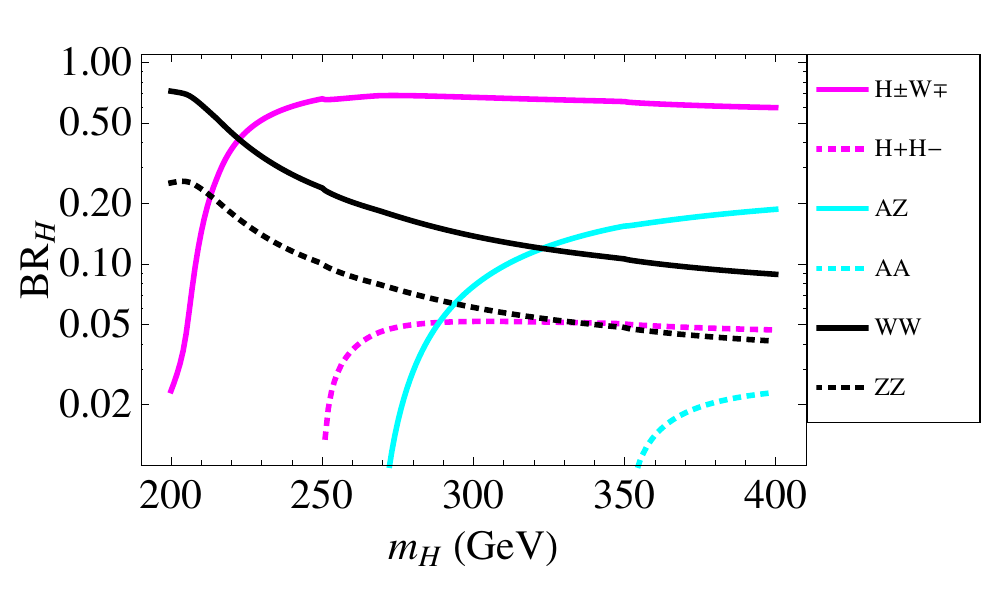}
\end{center}
\caption{Branching ratios of the $H^0$ boson as a function of its mass. Parameter values are
$m_h=125$~GeV, $m_\hp=125$~GeV, $m_A=175$~GeV, $\lambda_3 = 0$, $\sina = 0.9$, $\lambda_2 = \lambda_1$ and $ \lambda_7 = \lambda_6 $.
\label{fig:Hdecay1}}
\end{figure}

The upshot is that to obtain an absolute upper limit on $\sigma\times\text{BR}$ in our model,
we need only rescale the SM result by the factor $\ca^2$. As an example, in
 \reffig{fig:constraints}a, we have chosen $\sa=0.9$, so that $\ca^2=0.19$.
 Referring to the SM curve in Figure~12a of ref.\ \cite{ATLAS-CONF-2013-013}, we
 see that rescaling this curve by a factor 0.19, we are below or close to the
 exclusion limit for all masses above 200~GeV. If we instead take $\sa=0.8$,
our conservative estimate is closer to the exclusion region. 
Note that this is the upper limit
 in our model, and in most of parameter space one or more of the additional
 decay channels will be open, leading to a smaller $\BR(H^0\to ZZ)$. We conclude
 that at least for $\sa=0.9$ or larger, a $H^0$ in the mass range 200--500~GeV
 would not have been discovered at the LHC, and this also holds for smaller
 $\sa$ if additional decay channels are open. In this case, the limits become
 model dependent, and we leave this for future studies. 
As a first illustration of possible additional decay channels we calculate
the branching ratios for $H^0$ for one point in parameter space, as shown in
\reffig{fig:Hdecay1}. It is clear that as soon as the $H^0\to H^\pm W^\mp$ and/or $A^0 Z$ 
channels open, they will quickly dominate over the $WW$ and $ZZ$ channels, making $H^0$ 
more difficult to detect.

 We are now in a position to consider the actually observed Higgs boson ${\cal H}$ in our
 model. Similar studies have been done of the Inert Doublet Model in refs.~\cite{Swiezewska:2012eh,Goudelis:2013uca}.
 
 The most significant channel observed is the $\gamma\gamma$ decay, where
 the ATLAS experiment has observed a slight excess in signal strength, $\mu=1.65
 \pm 0.24\text{(stat)}^{+0.25}_{-0.18} \text{(syst)}$, where $\mu$ is the
 observed signal strength relative to the SM, at a mass of
 $126.8$~GeV~\cite{ATLAS-CONF-2013-012}. The overall best fit mass from all
 channels is $125.5$~GeV. The CMS experiment, on the other hand, sees no excess, and has a
combined best fit mass of $125.8$~GeV \cite{CMS-PAS-HIG-13-001}.

\begin{figure}[tb]
\begin{centering}
\begin{tabular}{cc}
\includegraphics[width=0.5\columnwidth]{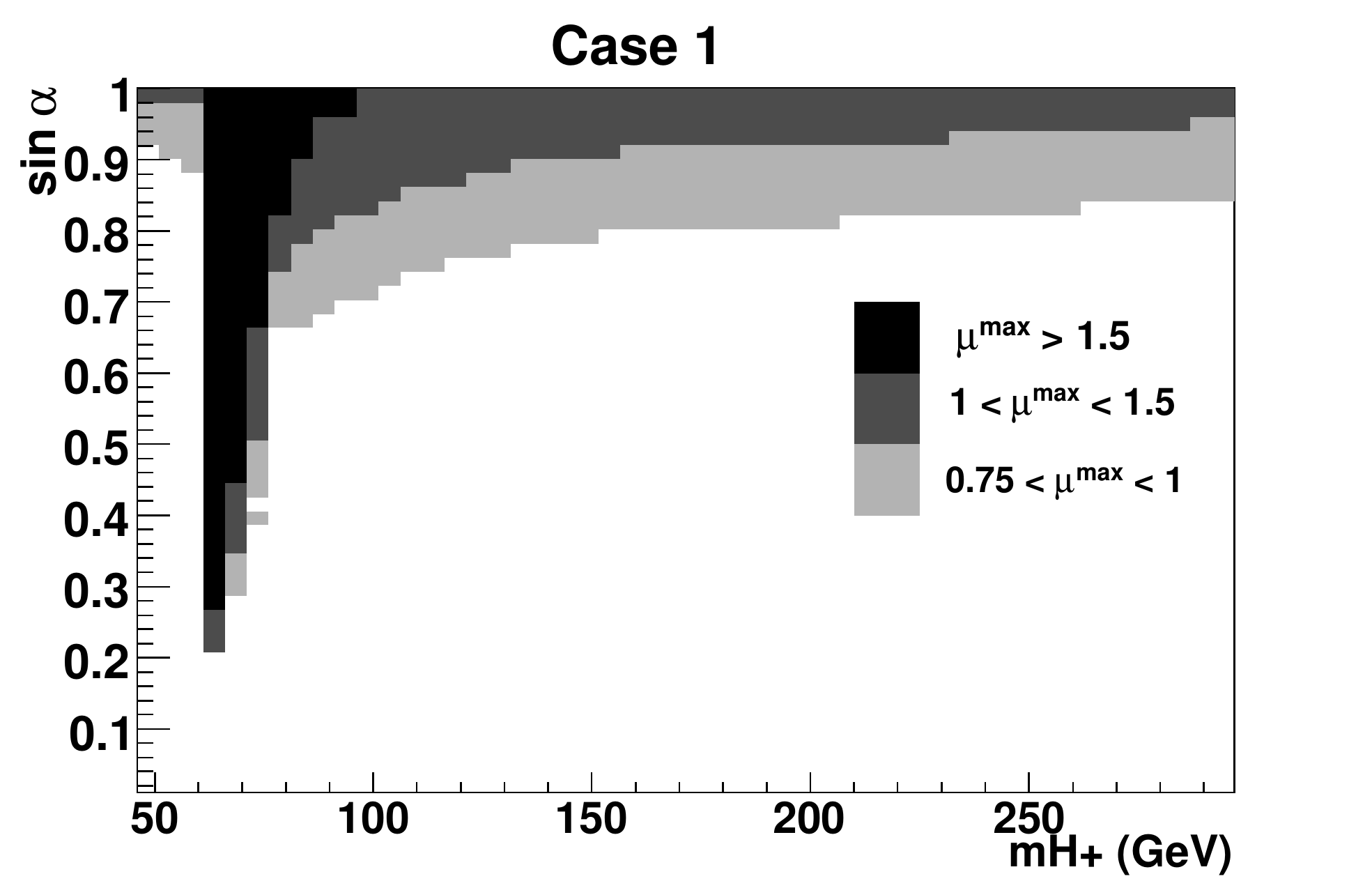}  &    
\includegraphics[width=0.5\columnwidth]{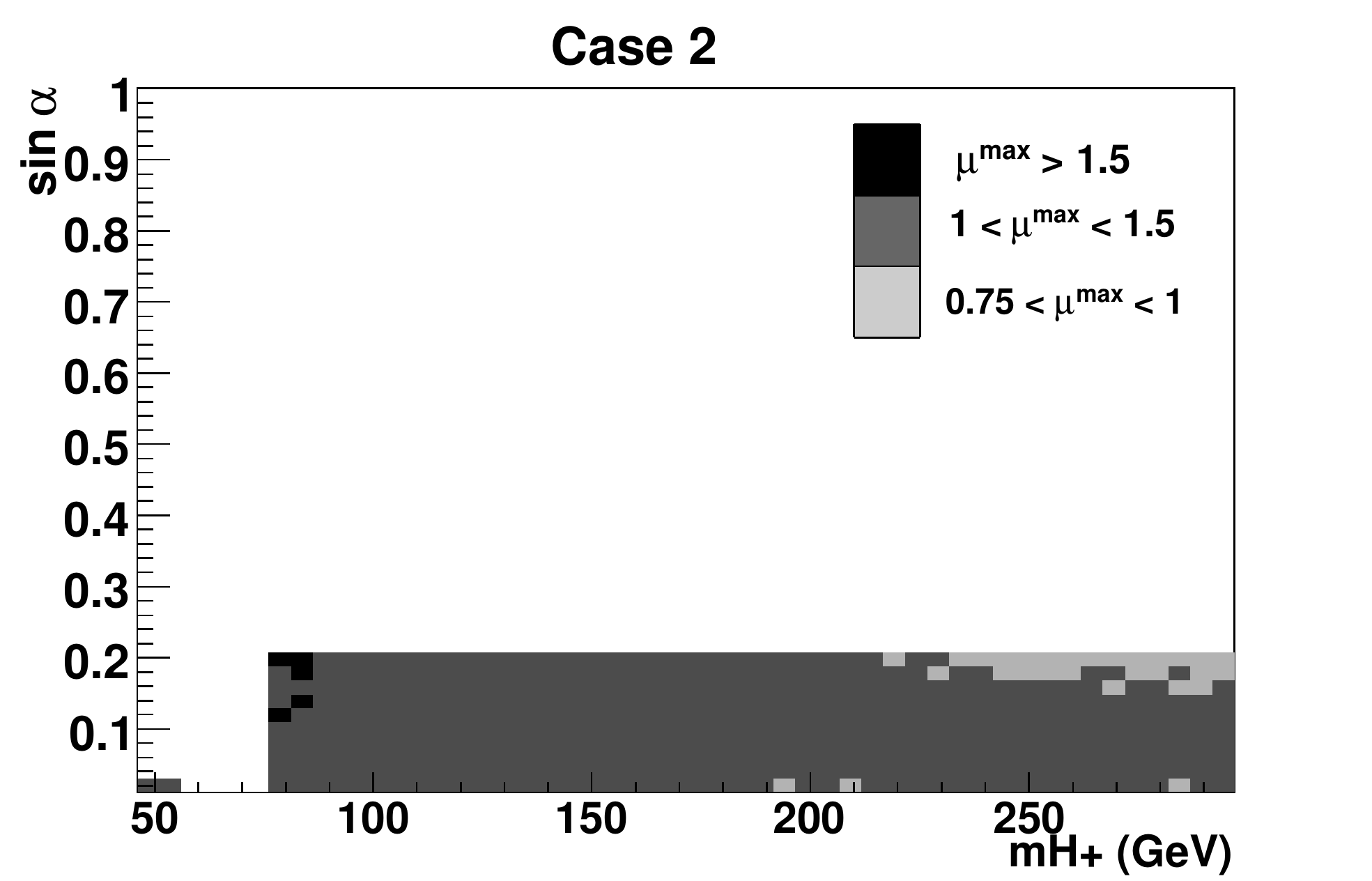} \\
$m_h = 125$ GeV,  $m_H = 300$ GeV &  $m_h = 75$ GeV, $m_H = 125$ GeV\\
(a) & (b)
\end{tabular}
\caption{The maximum $\mu_{\h\gamma\gamma}$ as described in the text, for (a) Case~1 and (b) Case 2, including
experimental and theoretical constraints.}
\label{fig:collider3}
\end{centering}
\end{figure}

 The signal strength for $\h\to\gamma\gamma$ with $\h=h^0,H^0$, relative to the
 SM, is computed as
\begin{align}
\mu_{\h\gamma\gamma} = \frac{\sum_i\sigma_i(pp\to \h) \BR(\h\to\gamma\gamma)}{\sum_i{\sigma}_i(pp\to H_\text{SM}) \BR(H_\text{SM}\to\gamma\gamma)}
\label{eq:signalstrength}
\end{align}
where the sums run over all contributing production channels. The production cross sections for $gg\to\h$, vector-boson
fusion, and associated production with a vector boson are all scaled in the same way compared to
the SM: there is a suppression factor $\sa^2$ for $h^0$ and $\ca^2$ for $H^0$. The signal strength for $\h=h^0$ can then be written
\begin{align}
\mu_{h^0\gamma\gamma} = \sa^2 \, \frac{\BR(h^0\to\gamma\gamma)}{\BR(H_\text{SM}\to\gamma\gamma)},
\end{align}
and $\mu_{H^0\gamma\gamma}$ is obtained in the same way, but replacing $\sa^2\to\ca^2$.
Unlike the case for the tree-level decays of the $H^0$ discussed above, the branching ratios here are
not equal in the SM and in our model, as there are charged scalars running in the loops.
We have performed a scan over $\lambda_3$ and $\lambda_7$, with $-5\le \lambda_3 \le 5$ and
$-5 \le \lambda_7 \le 5$, and in \reffig{fig:collider3} plot the maximum obtained signal strength provided that the theoretical, LEP, Tevatron and 7 TeV LHC constraints are fulfilled.\footnote{Note added:\ see figs 4 and 5a in \protect\cite{upcoming} for versions of these plots taking the latest LHC data into account.}
 We plot this
in the $(\mhp, \sa)$ plane for Cases 1 and 2 discussed above, fixing $m_{H}=300$~GeV for Case~1 and $m_{h}=75$~GeV for Case~2. The remaining parameter to be fixed is $m_A$.
In order to fulfill the EWPT constraints discussed above, we take, for Case~1, $m_A=\mhp + 50$~GeV if $m^2_\hp < M^2$ or $m_A=m_\hp$ if $m^2_\hp > M^2$, respectively (where $M^2$ was defined above), and for Case~2, we fix $m_A=\mhp$. If $m_h$ in Case~2 is increased, the sharp limit at $\sa\sim 0.2$ is shifted upwards.
We have not included off-shell decays $\h\to H^{+(*)} H^{-(*)}$ or $A^{0(*)} A^{0(*)}$ since the widths of the $H^\pm$ and $A^0$ are very small.

From \reffig{fig:collider3} we see that in order to have a maximal signal strength of one or larger, in Case~1, we need $\sa\gtrsim 0.8$, and in Case~2 we need $\sa\lesssim 0.2$. The enhancements are driven by light $\hp$ in the loop, with couplings to the CP-even scalars $g_{hH^+H^-} = -\ii v (-s_\alpha \lambda_3+c_\alpha \lambda_7)$ and
$g_{HH^+H^-} = -\ii v ( c_\alpha \lambda_3+s_\alpha \lambda_7)$.
If the charged scalar in Case~1 is very light ($\lesssim 80$~GeV), then it is possible to have almost any value for the mixing down to $\sa\sim 0.2$ in order to obtain the observed $\gamma\gamma$ signal strength. However, as discussed above, the signal strengths for $\h\to VV, f\bar f$ are given by 
$\mu_{h^0VV}=\mu_{h^0f\bar f}=\sa^2$ and $\mu_{H^0VV}=\mu_{H^0f\bar f}=\ca^2$ if $m_\hp>m_\h/2$. Therefore, from the observations of $\h\to ZZ$ and $\h\to WW$ at LHC, the low $\sa$ region in  \reffig{fig:collider3}a is disfavored.

Let us now comment on the phenomenology of the charged scalar. LEP has set limits on the mass: the model-independent limit from the $Z$ width is $m_{\hp}>39.6$~GeV~\cite{Abdallah:2003wd}, and there is a limit of $m_{\hp}\gtrsim 80$~GeV~\cite{Abdallah:2003wd,Heister:2002ev,Abbiendi:2013hk} under the assumption that $\BR(H^+\to\tau^+\nu)+\BR(H^+\to c\bar s)=1$. 
The charged scalar in our model has no tree-level fermion couplings, and does not primarily decay into $\tau\nu$ or $tb$ as in the MSSM. The available tree-level couplings are to gauge bosons and scalars, but the decays $\hp\to W^\pm\gamma$ and $\hp\to W^\pm Z$ are not allowed at the tree-level due to gauge and isospin invariance, respectively. The only possible tree-level decays are therefore to final states including at least one other scalar, which can be possibly off-shell leading to many-body final states. The usual decay channels $\tau\nu$ and $tb$ can indeed be generated at the one-loop level, but then we must also consider final states such as $W^\pm\gamma$  and $W^\pm Z$. In \cite{upcoming} we will perform a detailed study of these issues, but we note here that only the model independent LEP bound of 39.6~GeV applies, and that the $\hp\to W^\pm\gamma$ decay plays an important role, leading to a very interesting phenomenology.

\section{Conclusions}
In the Stealth Doublet Model presented in this paper, the $A^0$ and $\hp$ have no tree-level Yukawa couplings to fermions, so all such couplings are loop-suppressed. Flavor and LEP constraints therefore do not apply, and $A^0$ and $\hp$ can be lighter than in standard 2HDMs. 

The CP-even scalars $h^0$ and $H^0$ on the other hand, have very similar couplings to standard 2HDMs. We find that either the $h^0$ or the $H^0$ can play the role of the Higgs boson that has been observed at the LHC, and refer to these cases as Case~1 for $h^0$ and Case~2 for $H^0$. In both cases there are large allowed regions of parameter space where the observed $\h\to\gamma\gamma$ signal strength is obtained.

The phenomenology of the model depends on the mixing of the CP-even scalars, given by the parameter $\sa$, which also controls the amount of breaking of the $\ztwo$ symmetry; if $\sa\to 0$ or 1, we recover the Inert Doublet Model. 
In the two cases considered, when the observed $\h=h^0$, the data imply $\sa\gtrsim 0.8$, and when $\h=H^0$, $\sa\lesssim 0.2$ (valid for $m_h=75$~GeV). 
The masses of the charged and CP-odd scalars are constrained from the oblique parameters ($m_\hp\approx m_A$ or $m_\hp^2 \approx m_h^2\ca^2+m_H^2\sa^2$) and tree-level unitarity
($m_\hp, m_A \lesssim 700$~GeV). Moreover, $m_\hp \gtrsim m_\h/2$ is needed to obtain the observed $\gamma\gamma$ signal strength.

It is interesting to note that the parameter space that allows a signal strength for $\h\to\gamma\gamma$ of the same magnitude as the one observed at the LHC, is also the same parameter space
($\sa\gtrsim 0.8$) where the $H^0$ of Case~1 is safe from having been excluded. If Case~1 of our model is realized in Nature, we can therefore expect a discovery of the $H^0$ boson in the $H^0\to WW, ZZ$, $H^\pm W^\mp$ or $A^0 Z$ decay channels with more collected luminosity. If Case~2 is realized, we must instead look for a lighter $h^0$ boson. We will return to these issues and the study of the $\hp$ and $A^0$ in a forthcoming publication \cite{upcoming}.

\acknowledgments

We thank A. Arhrib, T. Hahn, and R. Pasechnik for helpful discussions. RE and JR are supported by the Swedish
Research Council under contracts 2007-4071 and 621-2011-5333.

\bibliographystyle{JHEP}
\bibliography{sdm}

\providecommand{\href}[2]{#2}\begingroup\raggedright\begin{thebibliography}{10}

\bibitem{ATLAShiggs}
{\bf ATLAS} Collaboration, G.~Aad et~al., {\it {Observation of a new particle
  in the search for the Standard Model Higgs boson with the ATLAS detector at
  the LHC}},  {\em Phys.Lett.} {\bf B716} (2012) 1--29,
  [\href{http://xxx.lanl.gov/abs/1207.7214}{{\tt arXiv:1207.7214}}].

\bibitem{CMShiggs}
{\bf CMS} Collaboration, S.~Chatrchyan et~al., {\it {Observation of a new boson
  at a mass of 125 GeV with the CMS experiment at the LHC}},  {\em Phys.Lett.}
  {\bf B716} (2012) 30--61, [\href{http://xxx.lanl.gov/abs/1207.7235}{{\tt
  arXiv:1207.7235}}].

\bibitem{ATLAS-CONF-2013-012}
{\bf ATLAS} Collaboration, {\it {Measurements of the properties of the
  Higgs-like boson in the two photon decay channel with the ATLAS detector
  using 25 $fb^{-1}$ of proton-proton collision data}},  Tech. Rep.
  ATLAS-CONF-2013-012, CERN, Geneva, Mar, 2013.

\bibitem{ATLAS-CONF-2013-013}
{\bf ATLAS} Collaboration, {\it {Measurements of the properties of the
  Higgs-like boson in the four lepton decay channel with the ATLAS detector
  using 25 $fb^{-1}$ of proton-proton collision data}},  Tech. Rep.
  ATLAS-CONF-2013-013, CERN, Geneva, Mar, 2013.

\bibitem{CMS-PAS-HIG-13-001}
{\bf CMS} Collaboration, {\it {Updated measurements of the Higgs boson at 125
  GeV in the two photon decay channel}},  Tech. Rep. CMS-PAS-HIG-13-001, CERN,
  Geneva, 2013.

\bibitem{Branco:2011iw}
G.~Branco et~al., {\it {Theory and phenomenology of two-Higgs-doublet models}},
   {\em Phys.Rept.} {\bf 516} (2012) 1--102,
  [\href{http://xxx.lanl.gov/abs/1106.0034}{{\tt arXiv:1106.0034}}].

\bibitem{Deshpande:1977rw}
N.~G. Deshpande and E.~Ma, {\it {Pattern of Symmetry Breaking with Two Higgs
  Doublets}},  {\em Phys.Rev.} {\bf D18} (1978) 2574.

\bibitem{Ma:2006km}
E.~Ma, {\it {Verifiable radiative seesaw mechanism of neutrino mass and dark
  matter}},  {\em Phys.Rev.} {\bf D73} (2006) 077301,
  [\href{http://xxx.lanl.gov/abs/hep-ph/0601225}{{\tt hep-ph/0601225}}].

\bibitem{Barbieri:2006dq}
R.~Barbieri, L.~J. Hall, and V.~S. Rychkov, {\it {Improved naturalness with a
  heavy Higgs: An Alternative road to LHC physics}},  {\em Phys.Rev.} {\bf D74}
  (2006) 015007, [\href{http://xxx.lanl.gov/abs/hep-ph/0603188}{{\tt
  hep-ph/0603188}}].

\bibitem{upcoming}
R.~Enberg, J.~Rathsman, and G.~Wouda, {\it {Higgs phenomenology in the Stealth
  Doublet Model}},  \href{http://xxx.lanl.gov/abs/1311.4367}{{\tt
  arXiv:1311.4367}}.

\bibitem{Wouda:2010zz}
G.~Wouda, {\it {Charged scalars in a lopsided doublet model}},  {\em PoS} {\bf
  CHARGED2010} (2010) 032.

\bibitem{Davidson:2005cw}
S.~Davidson and H.~E. Haber, {\it {Basis-independent methods for the
  two-Higgs-doublet model}},  {\em Phys. Rev.} {\bf D72} (2005) 035004,
  [\href{http://xxx.lanl.gov/abs/hep-ph/0504050}{{\tt hep-ph/0504050}}]. {\it
  Erratum:} Phys. Rev. {\bf D72} 099902 (2005).

\bibitem{Haber:2006ue}
H.~E. Haber and D.~O'Neil, {\it {Basis-independent methods for the
  two-Higgs-doublet model. II: The significance of tan(beta)}},  {\em Phys.
  Rev.} {\bf D74} (2006) 015018,
  [\href{http://xxx.lanl.gov/abs/hep-ph/0602242}{{\tt hep-ph/0602242}}]. {\it
  Erratum:} Phys. Rev. {\bf D74} 059905(E) (2006).

\bibitem{Bijnens:2011gd}
J.~Bijnens, J.~Lu, and J.~Rathsman, {\it {Constraining General Two Higgs
  Doublet Models by the Evolution of Yukawa Couplings}},  {\em JHEP} {\bf 1205}
  (2012) 118, [\href{http://xxx.lanl.gov/abs/1111.5760}{{\tt
  arXiv:1111.5760}}].

\bibitem{Eriksson:2009ws}
D.~Eriksson, J.~Rathsman, and O.~St{\aa}l, {\it {2HDMC: Two-Higgs-Doublet Model
  Calculator Physics and Manual}},  {\em Comput.Phys.Commun.} {\bf 181} (2010)
  189--205, [\href{http://xxx.lanl.gov/abs/0902.0851}{{\tt arXiv:0902.0851}}].

\bibitem{Eriksson:2010zzb}
D.~Eriksson, J.~Rathsman, and O.~St{\aa}l, {\it {2HDMC: Two-Higgs-doublet model
  calculator}},  {\em Comput.Phys.Commun.} {\bf 181} (2010) 833--834.

\bibitem{Sher:1988mj}
M.~Sher, {\it {Electroweak Higgs Potentials and Vacuum Stability}},  {\em Phys.
  Rept.} {\bf 179} (1989) 273--418.

\bibitem{Huffel:1980sk}
H.~Huffel and G.~Pocsik, {\it {Unitarity bounds on Higgs boson masses in the
  Weinberg- Salam model with two higgs doublets}},  {\em Zeit. Phys.} {\bf C8}
  (1981) 13.

\bibitem{Maalampi:1991fb}
J.~Maalampi, J.~Sirkka, and I.~Vilja, {\it {Tree level unitarity and triviality
  bounds for two Higgs models}},  {\em Phys. Lett.} {\bf B265} (1991) 371--376.

\bibitem{Kanemura:1993hm}
S.~Kanemura, T.~Kubota, and E.~Takasugi, {\it {Lee-Quigg-Thacker bounds for
  Higgs boson masses in a two doublet model}},  {\em Phys. Lett.} {\bf B313}
  (1993) 155--160, [\href{http://xxx.lanl.gov/abs/hep-ph/9303263}{{\tt
  hep-ph/9303263}}].

\bibitem{Akeroyd:2000wc}
A.~G. Akeroyd, A.~Arhrib, and E.-M. Naimi, {\it {Note on tree-level unitarity
  in the general two Higgs doublet model}},  {\em Phys. Lett.} {\bf B490}
  (2000) 119--124, [\href{http://xxx.lanl.gov/abs/hep-ph/0006035}{{\tt
  hep-ph/0006035}}].

\bibitem{PhysRevD.86.010001}
{\bf Particle Data Group} Collaboration, J.~Beringer et~al., {\it {Review of
  Particle Physics (RPP)}},  {\em Phys.Rev.} {\bf D86} (2012) 010001.

\bibitem{Peskin:1991sw}
M.~E. Peskin and T.~Takeuchi, {\it {Estimation of oblique electroweak
  corrections}},  {\em Phys. Rev.} {\bf D46} (1992) 381--409.

\bibitem{Bechtle:2008jh}
P.~Bechtle et~al., {\it {HiggsBounds: Confronting Arbitrary Higgs Sectors with
  Exclusion Bounds from LEP and the Tevatron}},  {\em Comput. Phys. Commun.}
  {\bf 181} (2010) 138--167, [\href{http://xxx.lanl.gov/abs/0811.4169}{{\tt
  arXiv:0811.4169}}].

\bibitem{Bechtle:2011sb}
P.~Bechtle et~al., {\it {HiggsBounds 2.0.0: Confronting Neutral and Charged
  Higgs Sector Predictions with Exclusion Bounds from LEP and the Tevatron}},
  {\em Comput.~Phys.~Commun.} {\bf 182} (2011) 2605--2631,
  [\href{http://xxx.lanl.gov/abs/1102.1898}{{\tt arXiv:1102.1898}}].

\bibitem{Bechtle:2013gu}
P.~Bechtle et~al., {\it {Recent Developments in HiggsBounds and a Preview of
  HiggsSignals}},  {\em PoS} {\bf CHARGED2012} (2013) 024,
  [\href{http://xxx.lanl.gov/abs/1301.2345}{{\tt arXiv:1301.2345}}].

\bibitem{Mahmoudi:2009zx}
F.~Mahmoudi and O.~St{\aa}l, {\it {Flavor constraints on the two-Higgs-doublet
  model with general Yukawa couplings}},
  \href{http://xxx.lanl.gov/abs/0907.1791}{{\tt arXiv:0907.1791}}.

\bibitem{Swiezewska:2012eh}
B.~Swiezewska and M.~Krawczyk, {\it {Diphoton rate in the Inert Doublet Model
  with a 125 GeV Higgs boson}},  \href{http://xxx.lanl.gov/abs/1212.4100}{{\tt
  arXiv:1212.4100}}.

\bibitem{Goudelis:2013uca}
A.~Goudelis, B.~Herrmann, and O.~St\aa{}l, {\it {Dark matter in the Inert
  Doublet Model after the discovery of a Higgs-like boson at the LHC}},
  \href{http://xxx.lanl.gov/abs/1303.3010}{{\tt arXiv:1303.3010}}.

\bibitem{Abdallah:2003wd}
{\bf DELPHI Collaboration} Collaboration, J.~Abdallah et~al., {\it {Search for
  charged Higgs bosons at LEP in general two Higgs doublet models}},  {\em
  Eur.Phys.J.} {\bf C34} (2004) 399--418,
  [\href{http://xxx.lanl.gov/abs/hep-ex/0404012}{{\tt hep-ex/0404012}}].

\bibitem{Heister:2002ev}
{\bf ALEPH Collaboration} Collaboration, A.~Heister et~al., {\it {Search for
  charged Higgs bosons in $e^{+} e^{-}$ collisions at energies up to $\sqrt{s}$
  = 209-GeV}},  {\em Phys.Lett.} {\bf B543} (2002) 1--13,
  [\href{http://xxx.lanl.gov/abs/hep-ex/0207054}{{\tt hep-ex/0207054}}].

\bibitem{Abbiendi:2013hk}
{\bf ALEPH Collaboration, DELPHI Collaboration, L3 Collaboration, OPAL
  Collaboration, The LEP working group for Higgs boson searches} Collaboration,
  G.~Abbiendi et~al., {\it {Search for Charged Higgs bosons: Combined Results
  Using LEP Data}},  {\em Eur.Phys.J.C} (2013)
  [\href{http://xxx.lanl.gov/abs/1301.6065}{{\tt arXiv:1301.6065}}].

\end{thebibliography}\endgroup

\end{document}